\documentclass{pnastwo}

\usepackage[dvips]{graphicx}

\usepackage{amssymb,amsfonts,amsmath}

\contributor{Submitted to Proceedings
of the National Academy of Sciences of the United States of America}
\url{www.pnas.org/cgi/doi/10.1073/pnas.0709640104}
\copyrightyear{2008}
\issuedate{Issue Date}
\volume{Volume}
\issuenumber{Issue Number}

\newcommand{\CeOsAs}{CeOs$_{4}$As$_{12}$}

\newcommand{\LaOsAs}{LaOs$_{4}$As$_{12}$}

\newcommand{\LaFeP}{LaFe$_{4}$P$_{12}$}
\newcommand{\CeBiPt}{Ce$_3$Bi$_4$Pt$_3$}

\begin{document}



\title{The filled skutterudite CeOs$_{4}$As$_{12}$: a hybridization gap semiconductor}





\author{R.~E. Baumbach\affil{1}{Department of Physics and Institute for Pure and Applied Physical
Sciences, University of California at San Diego, La Jolla, CA
92093, USA}, P.~C. Ho\affil{2}{Department of Physics, California
State University, Fresno, Fresno, CA 93740,USA},T.~A.
Sayles\affil{1}{}, M.~B. Maple\affil{1}{},
R.~Wawryk\affil{3}{Institute of Low Temperature and Structure
Research, Polish Academy of Sciences, 50-950 Wroc{\l}aw, Poland},
T.~Cichorek\affil{3}{}, A.~Pietraszko\affil{3}{}, \and
Z.~Henkie\affil{3}{}}

\contributor{Submitted to Proceedings of the National Academy of Sciences
of the United States of America}

\maketitle

\begin{article}

\begin{abstract} -- X-ray diffraction, electrical resistivity, magnetization, specific
heat, and thermoelectric power measurements are presented for
single crystals of the new filled skutterudite compound {\CeOsAs},
which reveal phenomena that are associated with f - electron -
conduction electron hybridization. Valence fluctuations or Kondo
behavior dominates the physics down to $T$ $\sim$ 135 K. The
correlated electron behavior is manifested at low temperatures as
a hybridization gap insulating state. The small energy gap
$\Delta$$_1$/k$_B$ $\sim$ 73 K, taken from fits to electrical
resistivity data, correlates with the evolution of a weakly
magnetic or nonmagnetic ground state, which is evident in the
magnetization data below a coherence temperature $T$$_{coh}$
$\sim$ 45 K. Additionally, the low temperature electronic specific
heat coefficient is small, $\gamma$ $\sim$ 19 mJ/mol\ K$^2$. Some
results for the nonmagnetic analogue compound {\LaOsAs} are also
presented for comparison purposes. -- \end{abstract}






\section{Introduction}

The ternary transition metal pnictides with the chemical formula
$M$$T$$_4$$X$$_{12}$ ($M$ = alkali metal, alkaline earth,
lanthanide, actinide; $T$ = Fe, Ru, Os; $X$ = P, As, Sb) which
crystallize in the filled skutterudite structure (space group
\textbf{I}m$\overline{3}$) exhibit a wide variety of strongly
correlated electron phenomena.\cite{Maple03,Aoki05,Maple08,Sato08}
Many of these phenomena depend on hybridization between the rare
earth or actinide f - electron states and the conduction electron
states which, in some filled skutterudite systems, leads to the
emergence of semiconducting behavior. This trend is evident in the
cerium transition metal phosphide and antimonide filled
skutterudite systems, most of which are semiconductors where the
gap size is correlated with the lattice
constant.\cite{Sugawara05,Braun80,Braun80b} While these systems
have been studied in detail, the arsenide analogues have received
considerably less attention, probably due to materials
difficulties inherent to their synthesis. In this paper, we
present new measurements of electrical resistivity $\rho$,
magnetization $M$, specific heat $C$, and thermoelectric power $S$
which show that {\CeOsAs} has a nonmagnetic or weakly magnetic
semiconducting ground state with an energy gap $\Delta$ in the
range 45 K $\leq$ $\Delta$/k$_B$ $\leq$ 73 K and a small
electronic specific heat coefficient $\gamma$ $\sim$ 19 mJ/mol\
K$^2$. Thus, we suggest that {\CeOsAs} may be a member of the
class of compounds, commonly referred to as hybridization gap
semiconductors or, in more modern terms, Kondo insulators. In
these materials the localized f - electron states hybridize with
the conduction electron states to produce a small gap ($\sim$ 1 -
10 meV) in the electronic density of states. Depending on the
electron concentration, the Fermi level can be found within the
gap, yielding semiconducting behavior, or outside the gap, giving
metallic heavy fermion behavior. For a more complete discussion of
the topic, we refer the reader to several excellent
reviews.\cite{Fisk92,Riseborough00} For comparison purposes, we
also report results for the weakly magnetic analogue material
{\LaOsAs}.

\section{Experimental Details}

Single crystals of {\CeOsAs} and {\LaOsAs} were grown from
elements with purities $\geq$ 99.9 $\%$ by a molten metal flux
method at high temperatures and pressures, as reported
elsewhere.\cite{Henkie08} After removing the majority of the flux
by distillation, {\CeOsAs} single crystals with dimensions up to
$\sim$ 0.7 mm and {\LaOsAs} single crystals with dimensions up to
$\sim$ 0.3 mm were collected and cleaned in acid in an effort to
remove any impurity phases from the surfaces of the crystals. The
crystal structure of CeOs$_4$As$_{12}$ was determined by X-ray
diffraction on a crystal with dimensions of 0.28 $\times$  0.25
$\times$ 0.23 mm. A total of 5757 reflections (454 unique,
Rint=0.1205) were recorded and the structure was resolved by the
full matrix least squares method using the SHELX-97
program.\cite{Sheldrick85, Sheldrick87} A similar measurement was
made for the {\LaOsAs} crystals.

Electrical resistivity $\rho(T)$ measurements for temperatures $T$
$=$ 65 mK - 300 K and magnetic fields $H$ $=$ 0 - 9 T were
performed on the {\CeOsAs} crystals in a four-wire configuration
using a conventional $^4$He cryostat and a $^3$He - $^4$He
dilution refrigerator. Magnetization $M(T)$ measurements for $T$
$=$ 1.9 K - 300 K and $H$ $=$ 0 - 5.5 T were conducted using a
Quantum Design Magnetic Properties Measurement System on mosaics
of both {\CeOsAs} (m $=$ 154.2 mg) and {\LaOsAs} (m $=$ 138.9 mg)
crystals, which were mounted on a small Delrin disc using Duco
cement. Specific heat $C(T)$ measurements for $T$ $=$ 650~mK -
18~K and $H$ $=$ 0 and 5 T were made using a standard heat pulse
technique on a collection of 116 single crystals (m $=$ $65$~mg)
of {\CeOsAs} which were attached to a sapphire platform with a
small amount of Apiezon N grease in a $^{3}$He semiadiabatic
calorimeter. Thermoelectric power $S(T)$ measurements for $T$ $=$
0.5 K - 350 K for {\CeOsAs} single crystals was determined by a
method described elsewhere.\cite{Wawryk01}

\section{Results}
Single crystal structural refinement shows that the unit cells of
{\CeOsAs} and {\LaOsAs} have the {\LaFeP} - type structure
(\textbf{I}m$\overline{3}$ space group) with two formula units per
unit cell, and room temperature lattice constants $a$ = 8.519(2)
\AA ~ and $a$ = 8.542(1) \AA, respectively, in reasonable
agreement with earlier measurements of $a$ = 8.5249(3) \AA~and $a$
= 8.5296 \AA~ for {\CeOsAs} and $a$ = 8.5437(2) \AA~ for
{\LaOsAs}.\cite{Braun80,Sekine07} Other crystal structure
parameters for these compounds are summarized in Table I. The
displacement parameter $\emph{D}$ represents the average
displacement of an atom vibrating around its lattice position and
is equal to its mean-square displacement along the Cartesian axes.
The displacement parameters determined for {\CeOsAs} and {\LaOsAs}
are typical of the lanthanide filled
skutterudites.\cite{Sales97,Sales99} Table I also indicates that
the Ce sites in CeOs$_4$As$_{12}$ may not be fully occupied, as is
often the case for filled skutterudite materials. By comparison to
other lanthanide filled skutterudite arsenides, the lattice
constant of {\CeOsAs} is reduced from the expected lanthanide
contraction value, indicating that the Ce$^{3+}$ 4f electron
states are strongly hybridized with the conduction electron states
of the surrounding Os and As ions.

Electrical resistivity data for a typical {\CeOsAs} single crystal
specimen for $T$ $=$ 65 mK - 300 K and $H$ $=$ 0 and a
polycrystalline {\LaOsAs} reference specimen\cite{Shirotani00} are
shown in Fig. 1. For $H$ $=$ 0 T, $\rho(T)$ for {\CeOsAs}
decreases weakly with $T$ from a value near 1 m$\Omega$cm at 300 K
to 0.5 m$\Omega$cm at 135 K, below which $\rho(T)$ increases
continuously to low $T$. An extremely broad hump centered around
$\sim$ 15 K is superimposed on the increasing $\rho(T)$ which
saturates below 300 mK towards a constant value near 100
m$\Omega$cm. Above 135 K, $\rho(T)$ for {\CeOsAs} is similar to
that of polycrystalline {\LaOsAs}. However, below 135 K, {\LaOsAs}
continues to exhibit metallic behavior, wherein $\rho(T)$
saturates near 0.1 m$\Omega$cm before undergoing an abrupt drop at
the superconducting transition near 3.2 K. With the application of
a magnetic field (left Fig. 1 inset), $\rho(T)$ for {\CeOsAs}
increases for 3 K $\leq$ $T$ $\leq$ 300 K by a small and nearly
temperature independent amount and the broad hump moves towards
lower $T$. Below $\sim$ 3 K, $\rho(T)$ for {\CeOsAs} is
drastically affected by $H$, where the residual resistivity,
$\rho$$_0$, is reduced by more than a factor of 10 with $H$ = 5 T.
For fields greater than 5 T, the relative effect of increasing
magnetic field on $\rho(T)$ is less pronounced, as $\rho$$_0$
saturates near 6 m$\Omega$cm. Finally, it should be noted that the
residual resistivity for $H$ = 9 T is slightly larger than that
for $H$ = 7 T.

DC magnetization data, $\chi(T)$ = $M(T)/H$, collected in a small
constant field $H$ = 0.5 kOe, are shown in Fig. 2 for {\CeOsAs}
and {\LaOsAs}. For CeOs$_4$As$_{12}$, $\chi(T)$ shows a weak and
unusual $T$ dependence. Below 300 K, $\chi(T)$ decreases with
decreasing $T$, suggesting a maximum above 300 K. Below a first
minimum at $T$ $\sim$ 135 K, $\chi(T)$ increases with decreasing
$T$ down to $T$$_{\chi,0}$ $\sim$ 45 K, where it again decreases
with decreasing $T$ down to $\sim$ 20 K. Near 20 K, $\chi(T)$ show
a second minimum, below which a weak upturn persists down to 1.9 K
with a small feature between 2 and 4 K. The minimum in $\chi(T)$
near 135 K roughly corresponds to the temperature where $\rho(T)$
begins to increase with decreasing $T$. The low $T$ upturn in
$\chi(T)$ is weakly suppressed in field, as reflected in the low
temperature $M$ vs $H$ isotherms (not shown) which have weak
negative curvature, and only begin to show a tendency towards
saturation above $\sim$ 30 kOe. No hysteretic behavior is
observed. Similar measurements for {\LaOsAs} at $T$ = 1.9 K reveal
that the low $T$ magnetic isotherm is linear in field up to 55
kOe. The magnetic susceptibility of {\LaOsAs} is of comparable
magnitude to that of {\CeOsAs}, and shows a weak increase with
decreasing $T$ down to the superconducting transition temperature
$T$$_c$ $\sim$ 3.2 K, as is typical for many La based compounds
for which $\chi(T)$ often has a weak temperature
dependence.\cite{Wawryk08}

Displayed in Fig. 3 are specific heat divided by temperature
$C(T)/T$ vs $T$$^2$ data for $T$ $=$ 650~mK - 18~K in magnetic
fields $H$ = 0 and 5 T. For 650~mK $<$ $T$ $<$ 7~K, $C(T)/T$ is
described by the expression,
\begin{equation}\label{eqn6}
C(T)/T = \gamma + \beta T^2
\end{equation}
where $\gamma$ is the electronic specific heat coefficient and
$\beta$ $\propto$ $\theta$$_D$$^{-3}$ describes the lattice
contribution. Fits of eq. 1 to the zero field $C(T)/T$ data show
that $\gamma$($H$ = 0) $\sim$ 19 mJ/mol\ K$^2$ and
$\theta$$_D$($H$ = 0) $\sim$ 264 K. In contrast to $\rho(T)$, the
application of a 5 T magnetic field has only a modest affect on
$C(T)/T$, for which $\gamma$ remains constant while $\theta$$_D$
increases to $\sim$ 287 K. It should also be noted that a small
deviation from the linear behavior is seen at $T$$^2$ $\sim$ 40
K$^2$ ($T$ $\sim$ 6 K), as shown in the Fig. 3 inset.

The thermoelectric power $S(T)$ data for {\CeOsAs} and {\LaOsAs}
are shown in Fig. 4. Near 300 K, $S(T)$ for {\CeOsAs} is positive
and one order of magnitude larger than that for metallic
{\LaOsAs}, which is shown here for reference and studied in
greater detail elsewhere.\cite{Wawryk08} From room $T$, the
thermoelectric power for {\CeOsAs} increases with decreasing $T$
to a maximum value $S_{MAX}$ = 83 $\mu$V/K at $T_{MAX}$ = 135 K,
close to the $T$ where semiconducting behavior is first observed
in the $\rho(T)$ data. The $S(T)$ data then decrease with
decreasing $T$ down to a minimum $S_{min}$ = 31 $\mu$V/K at
$T$$_{min}$ = 12 K, close to the maximum of the broad hump in
$\rho(T)$. Additionally, there is a broad peak superimposed on the
decreasing $S(T)$ for $T$ $=$ 12 K - 60 K. Below 12 K, the $S(T)$
data go through a second maximum $S_{max}$ = 46 $\mu$V/K at
$T_{max}$ = 2 K and finally continue to decrease down to 0.5 K,
but remain positive.

\section{Discussion}
When viewed globally, measurements of x-ray diffraction,
$\rho(T)$, $M(T)$, $C(T)/T$, and $S(T)$ indicate that strong f -
electron - conduction electron hybridization dominates the physics
of {\CeOsAs}, yielding a variety of types of behavior which can be
roughly divided into high ($T$ $\geq$ 135 K) and low ($T$ $\leq$
135 K ) temperature regions. For $T$ $>$ 135 K, the unusual
behavior is most pronounced in the $\chi(T)$ data, which deviate
from the typical Curie - Weiss $T$- dependence expected for
Ce$^{3+}$ ions and instead decrease with decreasing $T$ to a broad
minimum near 135 K $\leq$ $T$ $\leq$ 150 K. Although it is
difficult to definitively determine the origin of this $T$-
dependence, possible mechanisms include valence fluctuations and
Kondo behavior. In the valence fluctuation picture, which was
introduced to account for the nonmagnetic behavior of $\alpha$ -
Ce,\cite{MacPherson71} SmS in its collapsed ``gold"
phase,\cite{Maple74} and SmB$_6$,\cite{Menth69} the 4f electron
shell of each Ce ion temporally fluctuates between the
configurations 4f$^1$ (Ce$^{3+}$) and 4f$^0$ (Ce$^{4+}$) at a
frequency $\omega$ $\approx$ k$_B$$T$$_{vf}$/$\hbar$, where
$T$$_{vf}$ separates magnetic behavior at high temperatures $T$
$\gg$ $T$$_{vf}$ and nonmagnetic behavior at low temperatures $T$
$\ll$ $T$$_{vf}$. In this phenomenological model, the decrease in
$\chi$($T$) with decreasing $T$ can be accounted for in terms of
an increase in the Ce ion valence $v(T)$ with decreasing $T$
towards 4+, corresponding to a decrease in the average 4f electron
shell occupation number $n$$_{4f}$($T$) towards zero. At $T$ $=$ 0
K, $\chi$($T$) should be approximately,

\begin{subequations}
\begin{equation}\label{eqn1}
\chi(0) \approx \frac{n_{4f^0}(T)\mu^2_{eff}(4f^0) +
n_{4f^1}(T)\mu^2_{eff}(4f^1)}{3k_B(T+T_{vf})}|_{T=0}
\end{equation}
\begin{equation}\label{eqn1}
= \frac{n_{4f^1}(0)\mu^2_{eff}(4f^1)}{3k_B(T_{vf})}
\end{equation}
\end{subequations}
since $\mu$$_{eff}$(4f$^0$) = 0. In this model, the value of
$T_{sf}$ can be estimated from the Curie - Weiss behavior of
$\chi(T)$ for $T$ $\gg$ $T_{sf}$ where it plays the role of the
Curie - Weiss temperature, and $n_{4f^1}$(0) can be inferred from
$\chi(0)$. Since the Curie - Weiss behavior would be well above
room $T$ for this case, it is not possible to estimate $T_{sf}$
and $n_{4f^1}$($T$). The behavior of $n_{4f^1}$(0) can also be
deduced from measurements of the lattice constant $a$ as a
function of $T$, by x - ray diffraction or thermal expansion
measurements. The increase in valence with decreasing $T$ is
reminiscent of the $\gamma$ - $\alpha$ transition in elemental Ce,
in which the valence undergoes a transition towards 4+ with
decreasing $T$ that is discontinuous (1st order) for pressures $P$
$<$ $P$$_c$ where $P$$_c$ is the critical point, and continuous
(2nd order) for $P$ $>$ $P$$_c$. According to this picture, it
would appear that the critical point for {\CeOsAs} is at a
negative pressure and the valence is increasing with decreasing
$T$. A great deal of effort has been expended to develop a
microscopic picture of the intermediate valence state, and the
reader is referred to two of many review articles on the
subject.\cite{Maple78,Jayaraman79}

In contrast to the intermediate valence picture, it is possible
that the Ce ions remain in the 3+ state at all $T$, and the
Ce$^{3+}$ magnetic moments are screened by the conduction electron
spins below a characteristic temperature $T_K$. This scenario is
commonly called the Kondo picture,\cite{Hewson93} and may be
appropriate for {\CeOsAs} if $T$$_K$ $>$ 300 K, leading to Curie -
Weiss behavior for $T$ $>$ $T_K$ and the observed $\chi(T)$
behavior for 135 K $<$ $T$ $<$ $T_K$. A rough approximation of
$T$$_K$ can be made using the expression,

\begin{equation}\label{eqn1}
T_{K} = N_A\mu_{eff}^2/3\chi_0k_B
\end{equation}
where $\mu_{eff}$ = 2.54$\mu_{B}$ is the Hund's rule value for
Ce$^{3+}$ ions and $\chi_0$ $\sim$ 1.2 $\times$
10$^{-3}$cm$^3$/mol is the saturated value of $\chi(T)$ at $T$
$\sim$ 135 K, representing the magnetic susceptibility of a
screened Kondo ion singlet.\cite{Gajewski98} Using this rough
expression, $T$$_K$ is found to be near 670 K, supporting the
notion of a high Kondo temperature. Moreover, if the Kondo
interpretation with a high $T_K$ is correct, the decrease of
$\rho(T)$ with decreasing $T$ may reflect the onset of Kondo
coherence above 300 K, in addition to the freezing out of phonons.
Finally, it is worth noting that the $\chi(T)$ data for {\LaOsAs}
indicate that spin fluctuations associated with the transition
metal atoms do not account for the unusual $\chi(T)$ behavior seen
for {\CeOsAs}.

The affects of f - electron - conduction electron hybridization
become even more pronounced below 135 K, where $\rho(T)$ evolves
into a gapped semiconducting state which persists to the lowest
$T$ measured. The increase in $\rho(T)$ is not well described by a
simple Arrhenius function, as would be expected for a
semiconductor with a single energy gap. Instead, the following
phenomenological expression,\cite{Mott61}

\begin{equation}\label{eqn1}
\rho^{-1}(T) = \displaystyle\sum_{i=0}^3 A_iexp(-\Delta_i/k_BT)
\end{equation}
where $A$$_1$ = 2.7 (m$\Omega$cm)$^{-1}$, $\Delta$$_1$/k$_B$ = 73
K, $A$$_2$ = 0.44 (m$\Omega$cm)$^{-1}$, $\Delta$$_2$/k$_B$ = 16 K,
$A$$_3$ = 0.18 (m$\Omega$cm)$^{-1}$ and $\Delta$$_3$/k$_B$ = 2.51
K, describes the data for 1 K $\leq$ $T$ $\leq$ 135 K, as shown in
the right Fig. 1 inset. A possible interpretation of this fit is
that the large gap $\Delta$$_1$ describes the intrinsic energy gap
for {\CeOsAs} while $\Delta$$_2$ and $\Delta$$_3$ describe
impurity donor or acceptor states in the gap (Kondo
holes),\cite{Sollie91a,Sollie91b,Schlottmann92} or, possibly, a
pseudogap. In this situation, it is naively expected that the
application of a magnetic field will act to close the energy gaps
($\Delta$$_1$, $\Delta$$_2$, $\Delta$$_3$) through the Zeeman
interaction,

\begin{equation}\label{eqn1}
\Delta_{n}(H) = \Delta_{n}(H=0) - g_J|J_z|\mu_BH
\end{equation}
where $\Delta_{n}(H=0)$ are the zero field gaps, $g$$_J$ = 6/7 is
the Land\'{e} $g$ factor for Ce$^{3+}$ ions, $J$$_z$ $=$ 5/2 is
the z component of the total angular momentum for Ce$^{3+}$ ions,
and $\mu_B$ is the Bohr magneton. According to this expression, it
is expected that a field $H$ $=$ $\Delta_{n}(H=0)$/$g_J |J_z|
\mu_B$ $\sim$ 1.74 T is sufficient to close the low $T$ gap. This
behavior is qualitatively shown in the left Fig. 1 inset where the
low $T$ resistivity is reduced by nearly 70 \% for $H$ $=$ 1.5 T.

Concurrent with the emergence of the semiconducting behavior,
$\chi(T)$ increases slowly down to $T$$_{\chi,0}$$\sim$ 45 K,
where it goes through a maximum and then continues to decrease
with decreasing $T$. Therefore, there is an apparent correlation
between the intrinsic energy gap value $\Delta_1$ $\sim$ 73 K
taken from fits to $\rho(T)$ and the broad maximum in $\chi(T)$
centered around $T$$_{\chi,0}$ $\sim$ 45 K. If $T$$_{\chi,0}$ is
taken to indicate the onset of a coherent nonmagnetic ground state
where $T$$_{\chi,0}$ $\sim$ $T$$_{coh}$ $\sim$ $\Delta_1$/k$_B$,
then the likely description of the ground state is the well known
many body hybridization gap picture, where the magnetic moments of
the Ce$^{3+}$ f - electron lattice are screened by conduction
electron spins via an antiferromagnetic exchange interaction.

In the context of this model, the weak upturn in $\chi(T)$ for $T$
$<$ 20 K is unexpected. For this reason, it is necessary to
consider whether it is intrinsic or extrinsic to {\CeOsAs}. As
shown in the inset to Fig. 2, various functions describe the $T$-
dependence of the upturn below $\sim$ 10 K, including both power
law and logarithmic functions of the forms,

\begin{equation}\label{eqn3}
\chi(T) = aT^{-m}
\end{equation}
and
\begin{equation}\label{eqn4}
\chi(T) = b - c~ln(T)
\end{equation}
where $a$ $=$ 1.96$\times$10$^{-3}$ cm$^3$K$^m$/mol and $m$ $=$
0.16, or $b$ $=$ 1.92$\times$10$^{-3}$ cm$^3$/mol and $c$ $=$
0.25$\times$10$^{-3}$ cm$^3$/mol, respectively. These weak power
law and logarithmic divergences in $T$ of physical properties are
often associated with materials that exhibit non-Fermi liquid
behavior due to the proximity of a quantum critical
point.\cite{Maple94,Maple95,Stewart01} On the other hand, it is
also possible that the upturn is the result of paramagnetic
impurities, as is often found for hybridization gap
semiconductors.\cite{Riseborough00,Hundley90} To address this
possibility, the low $T$ upturn in $\chi(T)$ was fitted using a
Curie - Weiss function of the form,

\begin{equation}\label{eqn1}
\chi(T) = \chi_0 + C_{imp}/(T-\Theta)
\end{equation}
where $\chi$$_0$ = 1.13 $\times$ 10$^{-3}$ cm$^3$/mol,
C$_{imp}$($T$) = 3.12 $\times$ 10$^{-3}$ cm$^3$K/mol, and $\Theta$
= -3 K. Although it is not possible, by means of this measurement,
to quantitatively determine the origin of such an impurity
contribution, a rough estimate of the impurity concentration can
be made by assuming that it is due to paramagnetic impurity ions,
such as rare earth ions like Ce$^{3+}$ or Gd$^{3+}$, which give
impurity concentrations near 0.4 \% and 0.04\% respectively.
Finally, to elucidate the significance of an impurity contribution
of this type, the hypothetical intrinsic susceptibility
$\chi$$_{int}$($T$) = $\chi(T)$ - $\chi$$_{imp}$($T$) is shown in
Fig. 2. The resulting $\chi$$_{int}$($T$) remains nearly identical
to $\chi(T)$ down to $\sim$ 100 K, where the affect of
$\chi$$_{imp}$($T$) becomes apparent. Near 55 K,
$\chi$$_{int}$($T$) goes through a weak maximum and then saturates
towards a value near 1.1 $\times$ 10$^{-3}$ cm$^3$/mol with
decreasing $T$, consistent with the onset of Kondo coherence
leading to a weakly magnetic or nonmagnetic ground state.

Additionally, the behavior of $C(T)/T$ at low $T$ supports the
hybridization gap interpretation, for which the low $T$ electronic
specific heat coefficient $\gamma$ $=$ $C/T$ is expected to be
zero in the fully gapped scenario, or small and finite if a low
concentration of donor or acceptor levels are present or if there
is a pseudogap.\cite{Schlottmann93} Similar results were seen for
the archetypal hybridization gap semiconductor {\CeBiPt}, for
which $\gamma$ $\sim$ 3.3 mJ/(mol\ Ce\ K$^2$). For hybridization
gap semiconductors, this value is expected to be sample dependent
and to correlate with the number of impurities in the
system.\cite{Riseborough00,Hundley94} Similar sample dependence
has been observed in both $\rho(T)$ and $C(T)/T$ measurements for
{\CeOsAs} specimens. Moreover, a comparison to other systems where
large effective masses are seen as the result of many body
enhancements can be made by computing an effective
Wilson-Sommerfeld ratio R$_W$ $=$
($\pi$$^2$k$_B$$^2$/3$\mu$$^2$$_{eff}$)($\chi$/$\gamma$). The
values $\chi$, $\chi$$_{int}$, and $\gamma$ are taken at 1.9 K to
probe the low temperature state. Also, the Hund's rule effective
magnetic moment $\mu$$_{eff}$ = 2.54 $\mu_B$ for Ce$^{3+}$ ions is
used. From these values, R$_W$ is calculated to be $\sim$ 1.04 and
$\sim$ 0.67, for the $\chi$ and $\chi$$_{int}$ cases. As expected,
these values are similar to those found in f-electron materials
where the low $T$ transport and thermodynamic properties are due
to conduction electrons with or without enhanced masses.

The $T$ dependence of the thermoelectric power provides further
insight into the hybridization gap picture. Since $S(T)$ remains
positive over the entire $T$ range, it appears that charge
carriers excited from the top of the valence band to the
conduction band dominate the transport behavior for all $T$.
However, a single energy gap and charge carrier picture does not
adequately describe the $T$ dependence of the data. Instead, there
are several unusual features which include, (1) $T_{MAX}$ = 135 K,
which corresponds to the high $T$ transition in $\rho$($T$) from
semimetallic to metallic behavior, (2) $T_{min}$ = 12 K, which is
close to the low $T$ portion of the extremely broad hump
superimposed on $\rho(T)$, and (3) $T_{max}$ = 2 K, which is close
to $\Delta$$_3$/k$_B$. Thus, there is a close correspondence
between the deviations from a single energy gap picture between
$S(T)$ and $\rho(T)$, although the deviations are more pronounced
in $S(T)$ since electrons and holes make contributions of opposite
sign to the total $S(T)$. To further characterize $S(T)$, it is
useful to consider the expression,\cite{Busch56}

\begin{equation}\label{eqn1}
S(T) =  - P(\mu_n, \mu_ p)(k_B/e)(\Delta/2k_B T) + Q
\end{equation}
which describes intrinsic effects in $S(T)$ for a gapped system,
such as activation of electrons from the lower band to the higher
one, which creates both hole and electron charge carriers. For
this expression, the coefficients $P$($\mu_n$, $\mu_ p$) $=$
($\mu$$_n$ - $\mu$$_p$)/($\mu$$_n$ + $\mu$$_p$) and $Q$ $=$ -
(k$_B$/$e$)[($a$/2k$_B$) - 2] - (3/4)ln($m$$_n$/$m$$_p$), are
defined by the electron mobility $\mu$$_n$, hole mobility
$\mu$$_p$, the temperature coefficient $a$ of $\Delta$ vs $T$, the
electron effective mass $m$$_n$, and the hole effective mass
$m$$_p$. Since the quantity $Q$($a$, $m$$_n$, $m$$_p$) is expected
to be weakly $T$ dependent, this expression is a useful tool for
identifying energy gap behavior, as was the case for the
semiconductor Th$_3$As$_4$.\cite{Henkie78} However, as shown in
the left Fig. 4 inset, there is only one $T$ range, 6.5 K $\leq$
$T$ $\leq$ 15 K, where this type of behavior is observed. A linear
fit to $S(T)$ in this $T$ range yields $\Delta$/k$_B$ = 2.2 K,
which is of the order of $\Delta$$_3$/k$_B$ found by fits to
$\rho(T)$. To further characterize $S(T)$, it is also of interest
to consider $dS/dT$ vs log$T$ (not shown), which was used
previously to analyze the compound
Ce$_x$Y$_{1-x}$Cu$_{2.05}$Si$_2$,\cite{Ocko01} where a change in
the slope was interpreted as evidence for a change of dominant
electronic transport mechanism. For {\CeOsAs}, inspection of
$dS/dT$ vs log$T$ yields a characteristic temperature $T$$^*$ = 32
K.

Behnia et al.\cite{Behnia04} have argued that, in the zero
temperature limit, the thermoelectric power should obey the
relation,

\begin{equation}\label{eqn1}
q = (S/T)(N_A e/\gamma)
\end{equation}
where $N$$_A$ is Avogadro's number, $e$ is the electron charge and
the constant $N$$_A$$e$ = 9.65 $\times$ 10$^4$ C\ mol$^{-1}$ is
the Faraday number. The dimensionless quantity $q$ corresponds to
the density of carriers per formula unit for the case of a free
electron gas with an energy independent relaxation time. Taking
the values $S/T$ $=$ 17.5 V/K$^2$ and $\gamma$ $=$ 19 mJ/mol\
K$^2$ at $T$ $=$ 0.5 K and 0.65 K, respectively, the quantity  $q$
$=$ 89 is calculated. This value is nearly two orders of magnitude
larger than that seen for most of the correlated electron metals
analyzed by Behnia et al., but is similar to that found for the
Kondo insulator CeNiSn ($q$ = 107), for which the Hall carrier
density was found to be 0.01/f.u.~at 5 K.\cite{Behnia04} Thus, the
$q$ determined for {\CeOsAs} is consistent with a related
hybridization gap insulator and may be indicative of a low charge
carrier density in {\CeOsAs} at low $T$.

\section{Summary}

The compound {\CeOsAs} is considered in the context of a simple
picture where valence fluctuations or Kondo behavior dominates the
physics down to $T$ $\sim$ 135 K. The correlated electron behavior
is manifested at low temperatures as a hybridization gap
insulating state, or using modern terminology, as a Kondo
insulating state. The size of the energy gap $\Delta$$_1$/k$_B$
$\sim$ 73 K, is deduced from fits to electrical resistivity data
and the evolution of a weakly magnetic or nonmagnetic ground
state, which is evident in the magnetization data below a
coherence temperature $T$$_{coh}$ $\sim$ 45 K. Additionally, the
low temperature electronic specific heat coefficient is small,
$\gamma$ $\sim$ 19 mJ/mol\ K$^2$, as is expected for hybridization
gap insulators. The thermoelectric power and electrical
resistivity data also indicate that either impurity states in the
energy gap or the presence of a pseudogap strongly perturbs the
physical behavior away from that of an ideal hybridization gap
insulator.





\begin{acknowledgments}
Research at the University of California, San Diego was supported
by the U. S. Department of Energy under grant no. DE
FG02-04ER46105 and the National Science Foundation under grant no.
NSF DMR0335173. Research at ILTSR, Wroc{\l}aw was supported by the
Polish Ministry of Science and Higher Education (Grant No. NN 202
4129 33).
\end{acknowledgments}





\end{article}









\begin{table}
\caption{Crystal structure data for {\LaOsAs} and {\CeOsAs}. The
atomic coordinates $\emph{x}$, $\emph{y}$, and $\emph{z}$ $=$ 0.25
for T atoms and 0 for M atoms. The listed quantities are the
following; $\emph{a}$ (\AA) is the unit cell parameter,
$\emph{x}$, $\emph{y}$, and $\emph{z}$ are the atomic coordinates
for the As atoms, $\emph{O}$ are the occupancy factors for the T,
As, and M atoms, $\emph{l}$$_{M,T:As}$ (\AA) are the chemical bond
lengths M - As and T - As, $\emph{D}$ (\AA$^2$ $\times$ 10$^3$)
are the displacement parameters for atoms T, As, M, and $R1/wR2$
are the final discrepancy factors in \%.}
\begin{tabular}{|c|c|c|}
\hline                           & {\LaOsAs}   & {\CeOsAs} \\
\hline $\emph{a}$                &  8.542(1)   & 8.519(2)  \\
\hline $\emph{x}$                & 0.1488      & 0.1485    \\
\hline $\emph{y}$                & 0.3491      & 0.3485    \\
\hline $\emph{z}$                & 0           & 0         \\
\hline $\emph{O}$$_T$            & 1.00        & 1.00      \\
\hline $\emph{O}$$_{As}$         & 1.00        & 1.00      \\
\hline $\emph{O}$$_M$            & 1.00        & 0.99      \\
\hline $\emph{l}$$_{M,As}$       & 3.2416      & 3.2350    \\
\hline $\emph{l}$$_{T,As}$       & 2.4544      & 2.4470    \\
\hline $\emph{D}$                & 6; 7; 14    & 3; 4; 12  \\
\hline $\emph{R1/wR2}$               & 3.49/8.55   & 4.02/11.0 \\
\hline
\end{tabular}
\end{table}

\begin{figure}
       \includegraphics[width=3.25in]{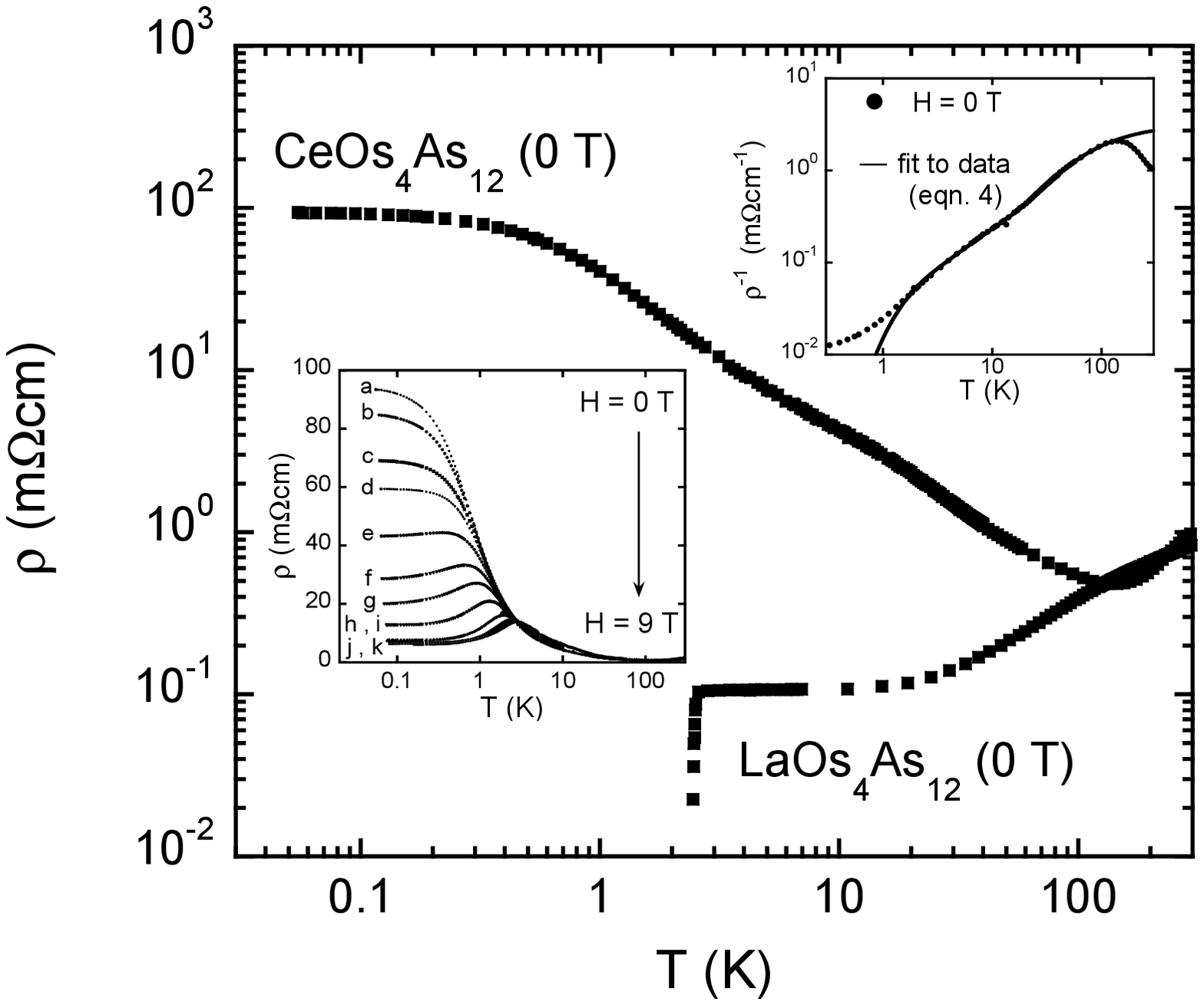}
\caption{Electrical resistivity $\rho$ vs temperature $T$ data for
{\LaOsAs}\cite{Shirotani00} and {\CeOsAs} for $T$ $=$ 65 mK - 300
K. Right inset close up: The zero field $\rho$($T$) data for
{\CeOsAs} are not described well by an Arrhenius equation which
would be appropriate for a simple single gap semiconductor.
Instead, the data are described by a three gap expression (eq. 4).
Left inset close up: $\rho(T)$ data for {\CeOsAs} for $T$ $=$ 65
mK - 300 K and magnetic fields (a) 0 T, (b) 0.3 T, (c) 0.5 T, (d)
0.7 T, (e) 1.0 T, (f) 1.5 T, (g) 2.0 T, (h) 3.0 T, (i) 5.0 T, (j)
7.0 T, and (k) 9.0 T.} \label{Fig1}
\end{figure}

\begin{figure}
       \includegraphics[width=3.25in]{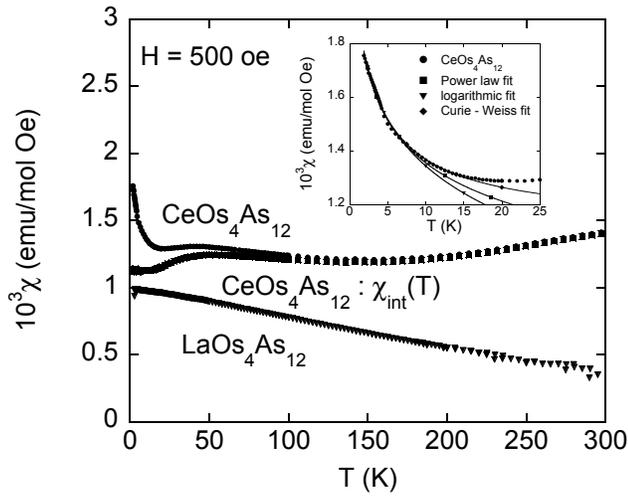}
\caption{Magnetic susceptibility $\chi$ = $M/H$ vs temperature $T$
data for {\CeOsAs} and {\LaOsAs} for $T$ $=$ 1.9 K - 300 K and $H$
= 500 Oe. Also shown is the intrinsic magnetic susceptibility
$\chi$$_{int}$($T$), as defined in the text, for which a Curie -
Weiss (CW) impurity contribution has been subtracted. Inset close
up: The low $T$ fits to $\chi(T)$ are described by eqs. 6, 7, and
8.} \label{Fig3}
\end{figure}

\begin{figure}
       \includegraphics[width=3.25in]{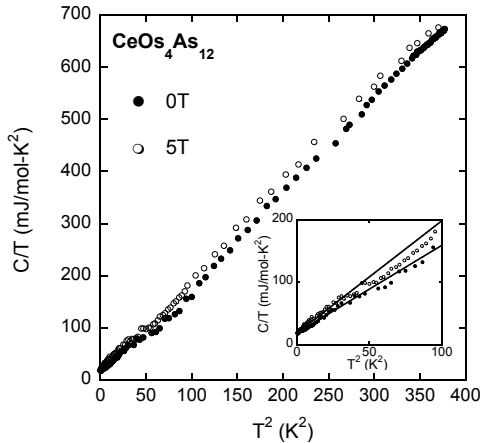}
\caption{Specific heat divided by temperature $C(T)/T$ vs
temperature squared $T^2$ data for {\CeOsAs} for $T$ $=$ 650~mK -
18~K and in magnetic fields of 0 and 5 T. Inset close up: $C(T)/T$
vs $T^2$ data for $T$ $=$ 650~mK - 10~K and in magnetic fields of
0 and 5 T. The straight lines are the fits to the data as
described in the text by eqn. 1.} \label{Fig4}
\end{figure}

\begin{figure}
       \includegraphics[width=3.25in]{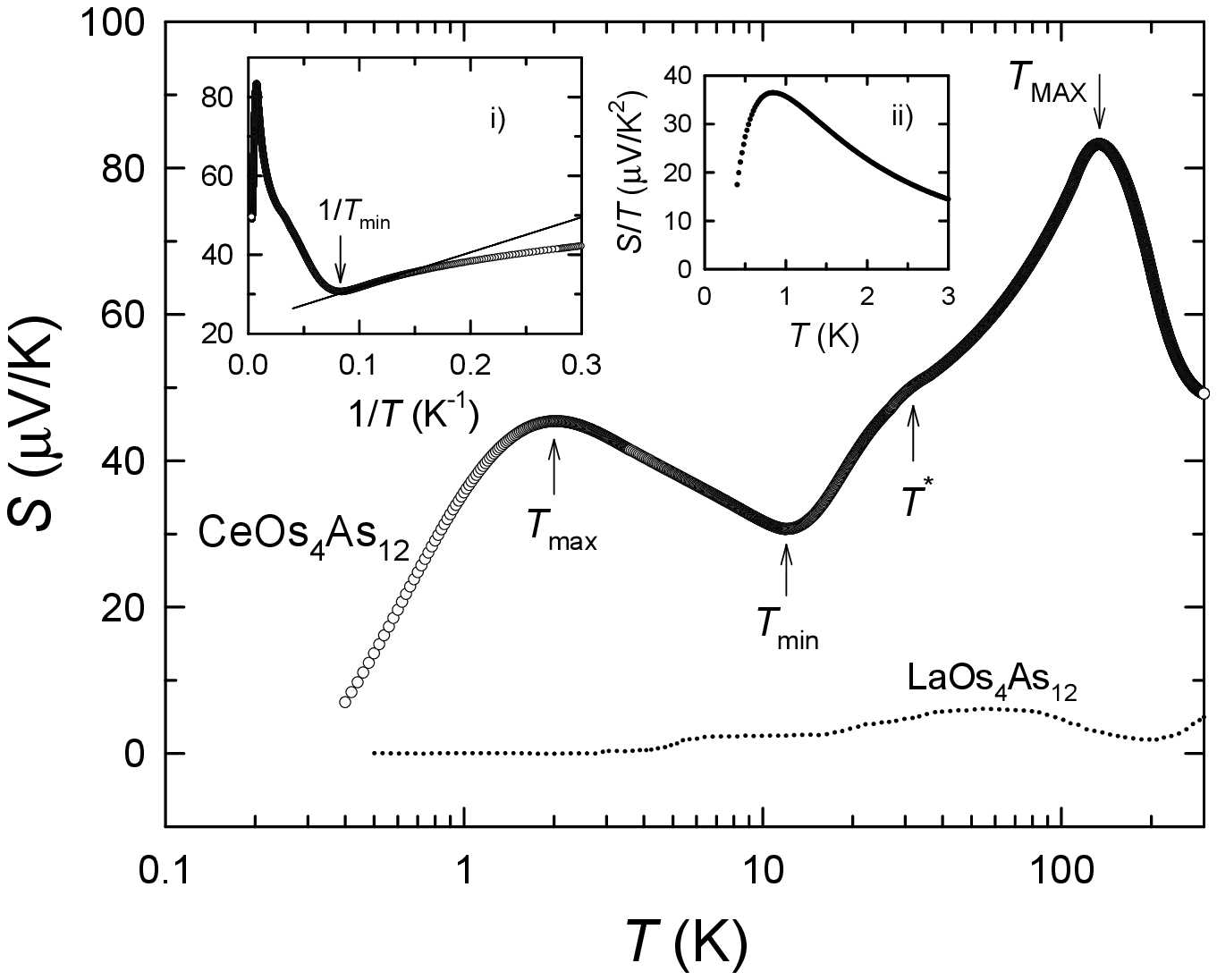}
\caption{Thermoelectric power $S$ vs temperature $T$ for {\CeOsAs}
and {\LaOsAs} for $T$ $=$ 0.5~K - 300~K. Near 300 K, $S(T)$ for
{\CeOsAs} is positive and one order of magnitude larger than that
for metallic {\LaOsAs}. Left inset close up: $S$ vs 1/$T$ is used
in the text to identify an energy gap. Right inset close up: $S/T$
vs $T$ is used to characterize the charge carrier density at low
$T$, as discussed in the text.} \label{Fig5}
\end{figure}

\end{document}